# Violations of Löwenstein's rule in zeolites

Rachel E. Fletcher, Sanliang Ling and Ben Slater*

Department of Chemistry, University College London, 20 Gordon Street, London
WC1H 0AJ, UK

*Email: b.slater@ucl.ac.uk

**Zeolites, microporous aluminosilicates, are amongst the most widely used catalysts in the petrochemical industry. Zeolite catalytic functionality is coupled to the distribution of tetrahedral alumina ($AlO_4^-$) and associated counter-cations throughout the aluminosilicate framework, yet little is definitively known about the factors that govern framework aluminium arrangement. It is generally accepted that all zeolites obey Löwenstein's rule of "aluminium avoidance", and that -Al-O-Al- formation is forbidden. Here, we describe the unprecedented screening of aluminium distribution in catalytically active zeolite SSZ-13 (CHA) in both its protonated and sodium-containing forms, H-SSZ-13 and Na-SSZ-13, using density functional theory (DFT). We predict violations of Löwenstein's rule in high and low silica H-SSZ-13 and other protonated frameworks considered in this investigation: H-LTA, H-RHO and H-ABW. The synthetic realisation of these zeolites could spur the development of new catalytic routes and materials, and the optimisation of existing zeolite catalysts.**

The use of zeolite catalysts in petrochemistry entirely revolutionised the industry over half a century ago. Since then, zeolites have become the workhorses of petrochemical processing, and are used extensively throughout the petrochemical industry.[1] Now, at a time of fast depleting traditional fuel sources and increasing toxic gas emission, zeolite catalysts are at the forefront of the development of 'green' alternatives to long-established petrochemical processes.[2] Green processes must operate at optimum efficiency[3] and for catalytic processes this requires the structural elucidation of existing catalytic materials. Unequivocally resolving a material's structure can expedite the identification of structure-activity relationships, which in turn, can accelerate the development of material specific design rules that are necessary for the rational design of new, more sustainable and efficient catalysts.

It is well understood that zeolite catalytic functionality originates from charged tetrahedral units of $AlO_4^-$ distributed throughout the aluminosilicate framework, and their associated charge-compensating cations located in nearby pores. Yet, despite major recent advances in experimental techniques,[4] at present it is not possible to determine the absolute position of framework aluminium or accompanying counter-cations exactly. Furthermore, there are currently no established design rules that can be applied to infer framework aluminium's preferred and precise position. However, Löwenstein's rule[5] of "aluminium avoidance" is commonly assumed; this states that on forming the aluminosilicate zeolite framework there is a disinclination for tetrahedral units of alumina to exist adjacent to one another, forbidding formation of -Al-O-Al- linkages, and restricting the maximum Si/Al ratio of any zeolite to unity.

Löwenstein's rule was conceptualised in 1954 and since then there have been few suggestions that violations of the rule are possible.[6–9] Indeed, the scientific literature reports that "aluminium avoidance" is observed in zeolites almost without exception.[5–11] Löwenstein's rule has hence become a fundamental law of zeolite science, and the possibility of non-Löwensteinian ordered zeolites are often dismissed. This is true of most theoretical studies where the omission of non-Löwensteinian frameworks is considered a simple way to reduce unnecessary computational expense by shrinking the number of potential configurations.[12,17–19] However, recent advances in supercomputing services and the development of increasingly efficient codes, mean it is now tractable to evaluate both Löwensteinian and non-Löwensteinian frameworks accurately through quantum mechanical approaches.

Characterisation techniques, such as X-ray diffraction, are currently unable to distinguish between framework silicon and catalytically important aluminium distributed throughout the zeolite. On the other hand, atomistic modelling techniques can be used as a tool to provide insight into the most probable location of framework aluminium in real zeolites. Using framework crystallographic data for a particular zeolite, quantum mechanical methods can unambiguously resolve the most energetically favourable distribution of both Si and Al. However, there is a further complication. It is well documented that the framework aluminium distribution of a given zeolite is highly dependent on the synthesis details.[20] The Si/Al ratio of the initial synthesis gel, synthesis temperature, reaction times, counter-cation identity and kinetic factors may all cause differences in the final structural chemistry, and hence catalytic activity of the resultant framework. Furthermore, Weckhuysen *et al.* recently showed that for a zeolite sample at a given Si/Al ratio, aluminium is inhomogenously distributed throughout the framework.[4,21] Commercial zeolites are typically synthesised using alkali metal cations as the charge compensating species, and facile ion-exchange techniques may be used post-synthesis to replace the metal cation with a proton, hence generating Brønsted acid O-H sites proximal to the location of aluminium. A key open question, which we partially address here, is whether the nature of the counter-cation affects the positioning of aluminium. However, the broader question is whether there is a clear thermodynamic incentive to form ordered or partially ordered frameworks and whether the position of aluminium can be predicted. Here, we present results obtained for the active small-pore zeolite catalyst SSZ-13, which displays a CHA-type framework topology[22]. The CHA framework (Figure 1) is made up of layers of hexagonally arranged double 6-ring (D6R) units connected by tilted 4-rings, giving rise to a characteristic *'cha'* cavity accessible through an 8-ring pore system.[22] Using periodic density functional theory (DFT) implemented in the program CP2K[23–25] we investigate all possible arrangements of framework aluminium, including non-Löwensteinian distributions, surveying the aluminium distribution of SSZ-13 at Si/Al ratios of 17, 11 and 8, in both Brønsted acidic H-SSZ-13 and the as-synthesised Na-SSZ-13. To our knowledge this is the only exhaustive study of zeolite framework aluminium distribution with different Si/Al ratios at this fully periodic quantum mechanical level of theory.

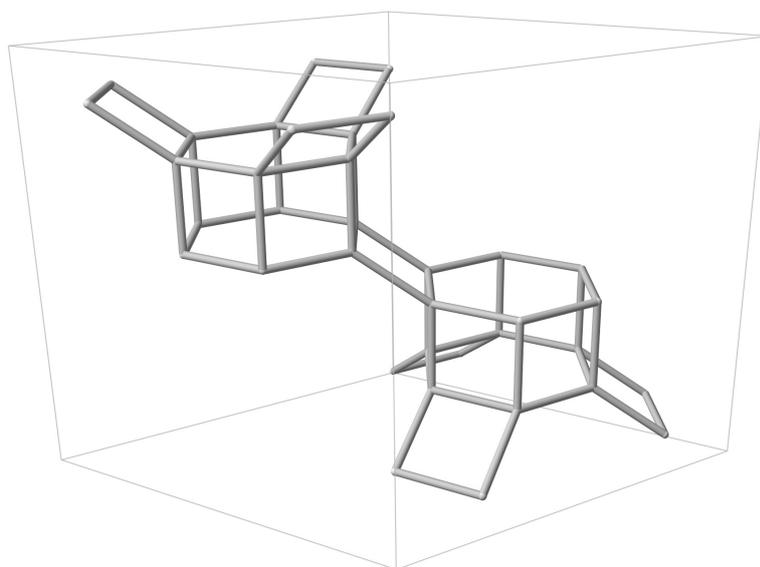

*Figure 1* Silicon backbone of the CHA framework contained within a single hexagonal unit cell, viewed along $[\bar{1}\ 1\ 0]$, displaying two D6R units and adjoining tilted 4-rings.

**Results**

A decisive variable in optimising catalytic activity is the Si/Al ratio as this dictates the density of charge compensating species, such as acidic sites. We compare three quite distinct Si/Al ratios to probe how the Si/Al ratio affects aluminium ordering.

**High silica SSZ-13 Si/Al = 17**

A single hexagonal CHA unit cell contains 36 crystallographically indistinct T-sites; in order to methodically explore all of the possible configurations of 2 Al per unit cell, corresponding to a Si/Al ratio of 17, a single aluminium atom, Al1, was substituted into an arbitrary T-site. Maintaining Al1's position, a second aluminium, Al2, was sequentially introduced into the remaining 35 T-sites. To maintain charge-neutrality, each individual aluminium substitution requires charge compensation by a cationic moiety. For H-SSZ-13, each cationic proton may reside at one of four oxygen sites at the apices of the alumina tetrahedral, yielding four potential topologically inequivalent Brønsted acid O-H sites per Al substitution, and hence a total of 560 unique combinations of 2 Al per unit cell. We used the periodic DFT method (at the PBE level)[26] to fully optimize each individual framework model to equilibrium density; the resulting data is shown in Figure 2, where the relative energy per unit cell (UC) (with respect to the average total energy) is given as a function of framework aluminium separation.

Assuming Löwenstein's rule[5] is valid, and the principle of aluminium avoidance is adhered to, we would expect the highest energy SSZ-13 structures to be those containing aluminium atoms at separations equivalent to that of a "forbidden" -Al-O-Al- linkage, and structures with larger aluminium separations to become increasingly more stable, in accordance with Dempsey's rule[27] (a less sophisticated rule which states, on the basis of electrostatics, that negatively charged alumina units are inclined to be positioned as far away from one another as possible).

As predicted from Löwenstein's rule, the highest energy configurations for both H-SSZ-13 and Na-SSZ-13 are those containing adjacent aluminium atoms, with a separation of approximately 3 Å (Figure 2). However, beyond this distance the relative energy landscapes for the two forms of the zeolite become dramatically different.

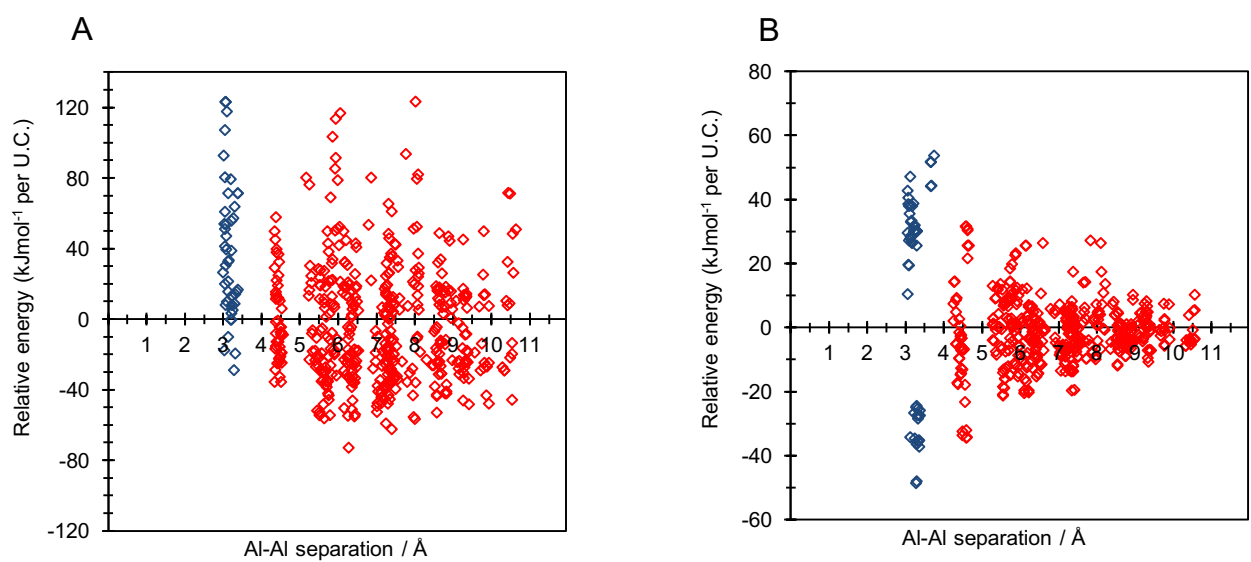

*Figure 2* Relative energy distribution (kJ mol$^{-1}$ per U.C.) against framework aluminium separation for *a)* Na-SSZ-13 and *b)* H-SSZ-13. Frameworks possessing non-Löwensteinian (NL) ordered aluminium atoms (-Al-O-Al-) are shown in blue.

In accordance with Löwenstein's rule, and what has already been widely observed in sodium-containing zeolites, the Na-SSZ-13 global minimum (Figure 3a) contains aluminium pairs as next-next nearest neighbours (NNNN),[12] and there is a +44·3 kJ mol$^{-1}$ per U.C. energy penalty for forming the most favourable non-Löwensteinian (NL) structure. Ignoring barriers, the penalty to form a NL structure is at least 10$k$T (assuming a typical synthesis temperature of ~450K), which suggests -Al-O-Al- linkages are very unfavourable in Na-SSZ-13. In the global minimum structure, the aluminium ions are separated by a distance of 6·18 Å, and their associated Na$^+$ cations reside at the parameters of the 8-ring apertures of the *'cha'* cavity. However, the H-SSZ-13 global minimum structure (Figure 3c) is remarkably different, containing adjacent aluminium ions along the edge of the 6-ring at a separation of 3·28 Å, violating Löwenstein's rule. In this structure the two associated protons, H1 and H2, are separated by 4·36 Å and arranged *trans* to one another; H1, which mediates the aluminium ions, is directed into the plane of the 6-ring, and H2, positioned at the connecting edge of the D6R, is oriented away from H1, and directed into the 8-ring window of the *'cha'* cavity. The most stable Löwensteinian (L) structure (Figure 3d) contains aluminium ions as next nearest neighbours (NNN), at a 'non-Dempsey' separation of 4·60 Å, with both protons directed into different 8-ring windows of the *'cha'* cavity. The energy penalty for forming the L structure is +14·2 kJmol$^{-1}$ per U.C. rather than the NL structure, is approximately one third of the energy difference between the global minimum NL/L structures for Na-SSZ-13. ΔE(NL$_{global\ minimum}$ − L$_{global\ minimum}$) for Na-SSZ-13 is +44·3 kJ mol$^{-1}$ per U.C., whilst ΔE(NL$_{global\ minimum}$ − L$_{global\ minimum}$) for H-SSZ-13 is -14·2 kJ mol$^{-1}$ per U.C., indicating a strong enthalpic incentive for Löwensteinian configurations when Na is the charge compensating cation and a modest enthalpic incentive to adopt non-Löwensteinian linkages when the charge compensator is a proton. Free energy calculations that include the vibrational entropy contributions to the energy show that the relative stability of the L and NL H-SSZ-13 configurations is maintained beyond typical synthesis temperatures (see SI) demonstrating a clear thermodynamic preference for adopting non-Löwensteinian structures for the proton compensated structure, a result that clearly conflicts with accepted wisdom. Furthermore, the NL H-SSZ-13 global minimum is not unique and seven other NL ordered frameworks (excluding the global minimum structure), all of which contain proton arrangements similar to those displayed in the NL global minimum structure (Figure 3c), are more stable than the global minimum L H-SSZ-13 structure.

To verify our unexpected H-SSZ-13 result we further investigated the lowest energy structures using the higher level hybrid functional PBE0 approach.[28,29] The results for these calculations are included in the supplementary information (S2), and show the relative energies for these structures to be essentially identical to those given by standard PBE, confirming the robustness of the predictions.

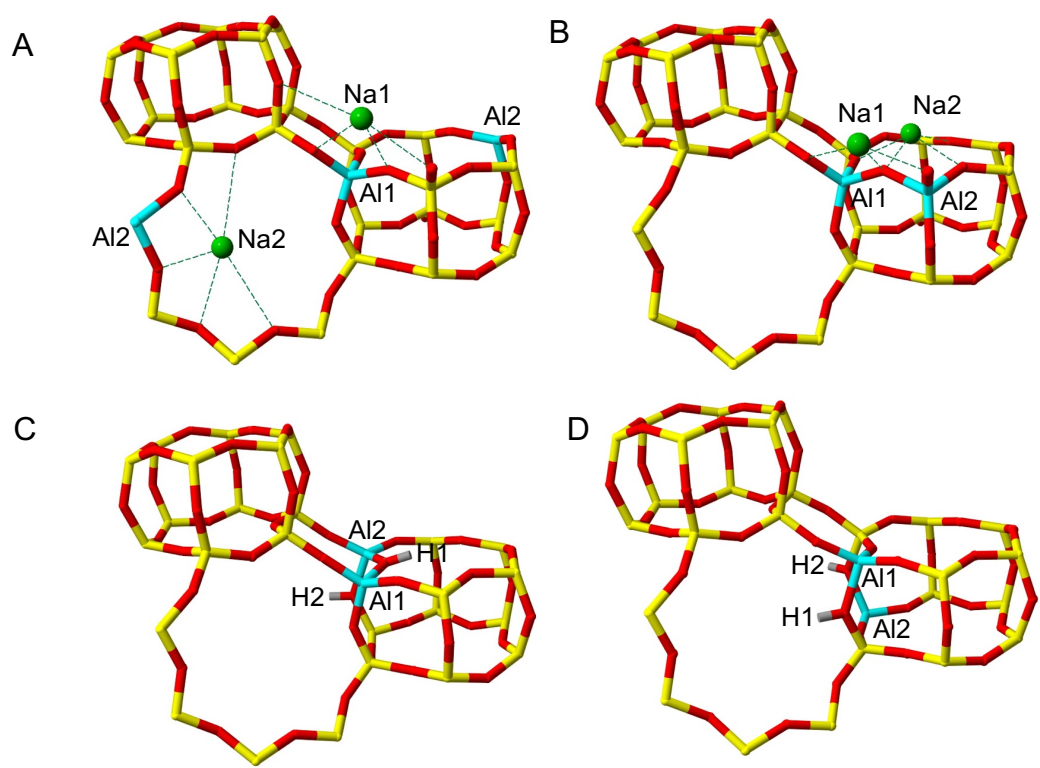

*Figure 3* Global minima L and NL 2 Al per U.C. SSZ-13 structures predicted by DFT. *a)* Global minimum Na-SSZ-13 structure, with Löwensteinian order aluminium atoms

*at the NNNN position, **b)** lowest energy NL Na-SSZ-13 structure, **c)** Global minimum NL H-SSZ-13 structure, with protons oriented trans to one another **d)** lowest energy Löwensteinian ordered structure with Al at NNN position. Where; silicon (yellow), oxygen (red), aluminium (blue), sodium (green), hydrogen (grey)*

**Low Silica SSZ-13 Si/Al < 17**

Exploring SSZ-13 with lower Si/Al ratios becomes increasingly complicated with each introduction of additional aluminium. To avoid calculating the prohibitively large number of combinations of 3 Al per unit cell SSZ-13 (Si/Al = 11), we employed a method of stepwise aluminium incorporation. In this approach, the NL and L ordered 2 Al per unit cell global minima according to the prior DFT (Figures 3c and d) were used as the initial structures. A single Al, Al3, was sequentially introduced into each of the remaining silica T-sites of both structures, and the appropriate counter-cation positioned at one of the four apical oxygen sites. A total of 136 distinct framework arrangements were created for each NL and L initial global minimum structure for Na and H (272 calculations for both Na and H).

Each structure was optimised and the NL and L H-SSZ-13 initial configurations gave the same 3 Al per unit cell global minimum structure. The structure, Figure 4a, contains a chain of three oxygen linked aluminium atoms, $[O-Al-O]_3$, with each charge-compensating proton located at a bridging oxygen and arranged *trans* to its neighbour(s). Once again, Na-SSZ-13 did not follow the same trend as H-SSZ-13, where each initial structure yielded different global minimum structures; the Löwensteinian structure containing the third aluminium at the next-nearest neighbour position,[15] and the non-Löwensteinian structure favouring Al at the NNNN position. The corresponding figures for these structures can be found in the SI.

Using the H-SSZ-13 (Si/Al = 11) global minimum as the new initial structure, we then proceeded to investigate 4 Al per unit cell, equivalent to a Si/Al ratio of 8. The global minimum structure, shown in Figure 4b, contains a chain of four oxygen linked aluminium atoms arranged in a 4-ring, with protons arranged *trans* to one another. In the sodium form of this structure (S5), the 4$^{th}$ Al resides in the next-nearest neighbour position (NNN), again in accordance with Löwenstein's rule, and what has already been documented for similar zeolites.[15] All four sodium cations position themselves proximal to the aluminium ions, at the centre of both faces of the aluminium doped D6R unit and at the parameters of the proximal 8-rings. It appears that as the aluminium content of the zeolite is increased, the aluminium clusters into zones of concentrated -Al-O-Al-, this is contrary to the general belief that that aluminium is reasonably well dispersed throughout the frameworks of real samples.[4,21]

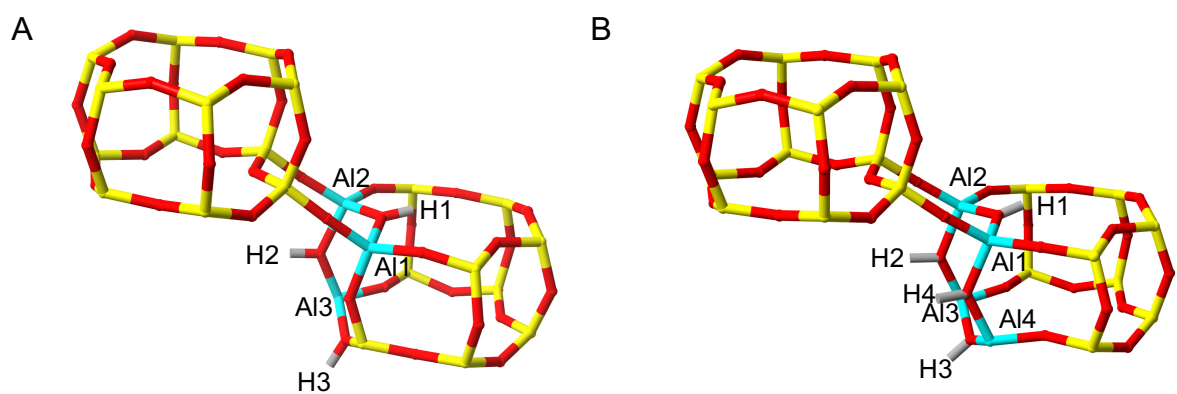

*Figure 4 a)* H-SSZ-13 3 Al per unit cell (Si/Al = 11) global minimum structure, containing a chain of 3 -Al-O- units. *b)* H-SSZ-13 4 Al per unit cell (Si/Al =8) global minimum structure, containing a chain of 4 -Al-O-Al- units

**Other zeolite framework types**

To ascertain whether our unusual findings are ubiquitous to all zeolite frameworks, or unique to CHA, we investigated a selection of framework-types by the same methods previously discussed. The selected frameworks, LTA, RHO and ABW, are shown in Figure 5, and their corresponding densities are included in the SI (S6).[22]

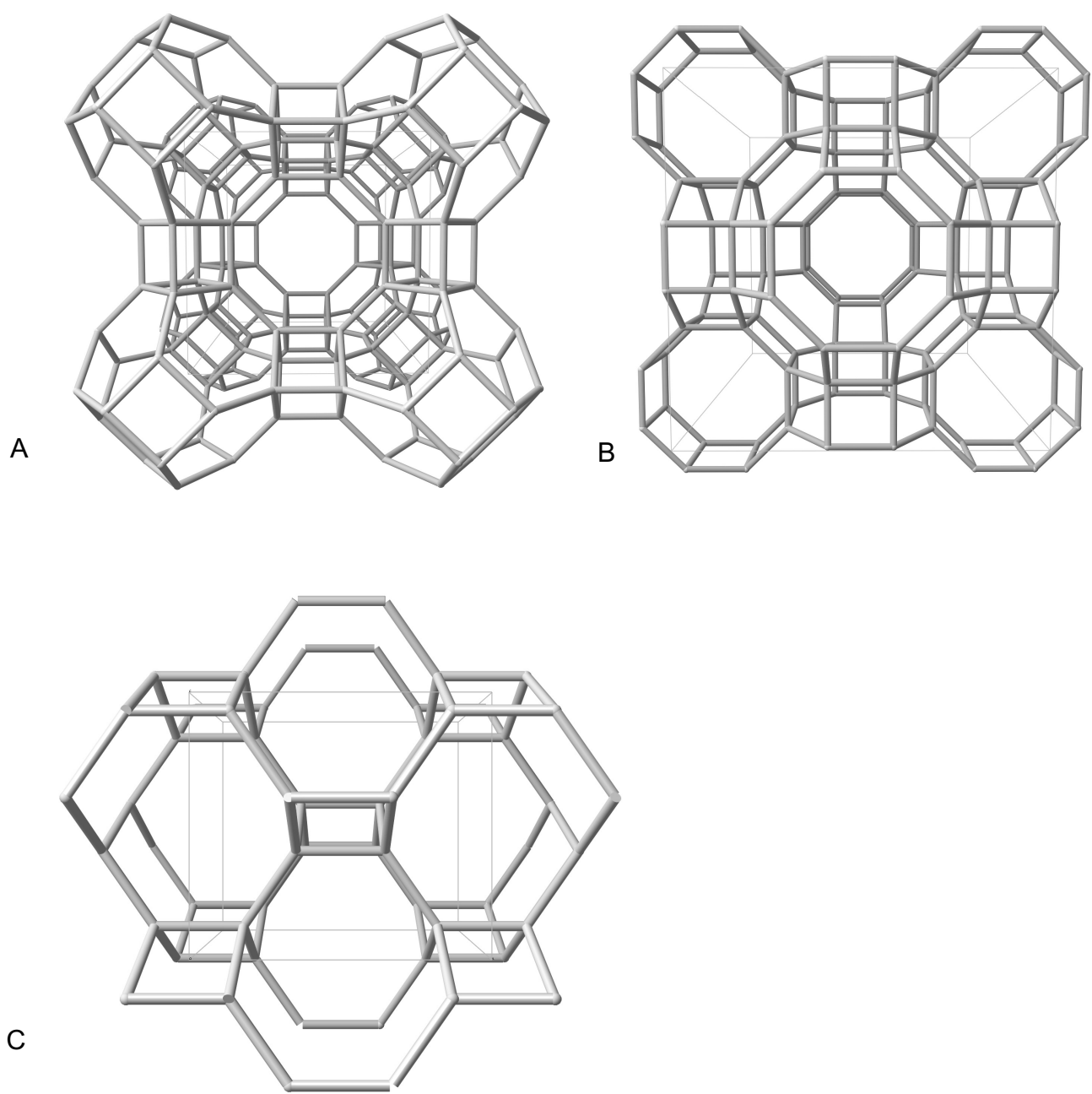

*Figure 5* Silicon backbones for the unit cells of additional zeolite frameworks considered in this work: *a)* LTA viewed along [1 0 0 ] *b)* RHO viewed along [1 0 0 ] *c)* ABW viewed along [ 0 1 0 ] [22]

Each of the frameworks contain a single crystallographically distinct T-site, and exhibit contrasting densities and topologies to that of CHA. The least dense of the frameworks, LTA, has a highly controversial history regarding aluminium distribution at Si/Al ratios tending to 1, where previous work has suggested the existence of non-Löwensteinian NL linkages.[6,7] Investigation of each framework (2 Al per unit cell) using DFT revealed that all three framework types possess NL global minimum structures in their protonated forms, and that the protons adopt the same *'trans'*-like orientation as seen for CHA. The data for each structure is shown in Figure 6, and the corresponding global minimum framework structures for each framework type can be found in the SI (S7). The energy penalty for forming the L structure ($\Delta E$ ($NL_{global\ minimum}$-$L_{global\ minimum}$)) for high density H-ABW is +55·7 kJ mol$^{-1}$ per U.C., +14.4 kJ mol$^{-1}$ per U.C. for H-CHA, +9·2 kJ mol$^{-1}$ per U.C. for H-RHO and +8.3 kJ mol$^{-1}$ per U.C. for H-LTA, which correlates with their respective densities. These results suggest that NL linkages are more strongly preferred in denser zeolites but even in LTA, which is one of the lowest density zeolites, the energy penalty for forming L structures is ~2$k$T at typical synthesis temperatures (~450K).

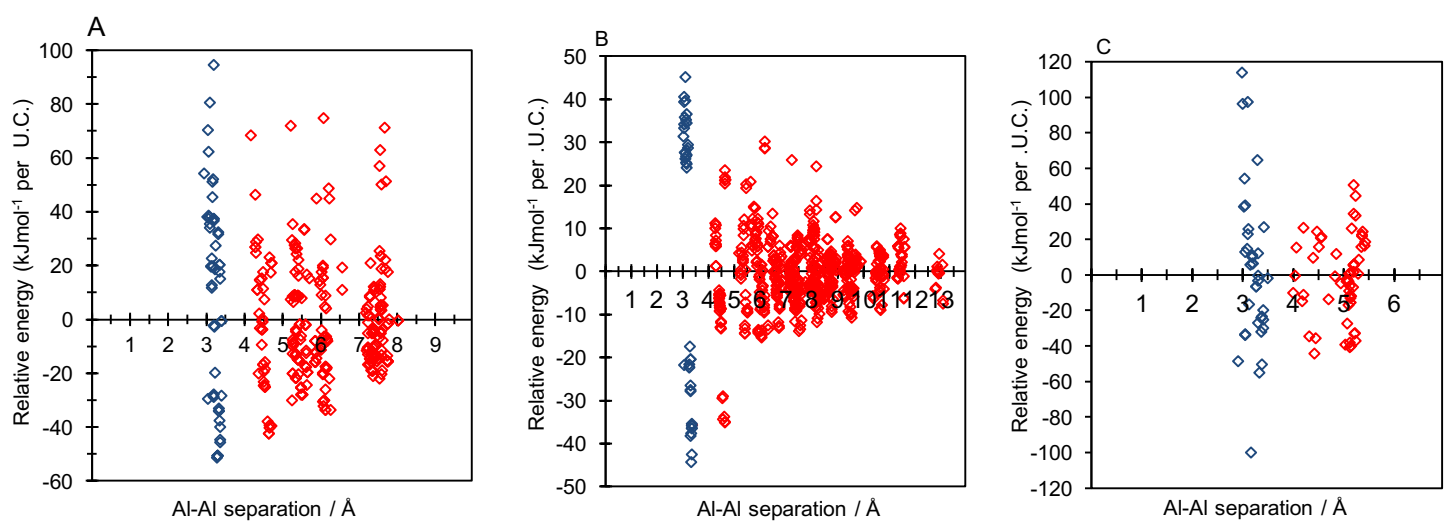

***Figure 6*** *Relative energy dispersion (kJ mol$^{-1}$ per UC) against framework aluminium separation for **a)** LTA (1:1:1) **b)** RHO (1:1:1) **c)** ABW (2:2:2). Frameworks possessing non-Löwensteinian (NL) ordered aluminium atoms (-Al-O-Al-) are shown in blue.*

**Discussion**

This work provides evidence for non-Löwensteinian ordering in protonated zeolite frameworks, where there is a thermodynamic preference for $Al^{3+}$ ions to exist adjacent to one another linked by a bridging hydroxyl moiety oriented *'trans'* to its nearest neighbour proton. This prediction holds true across a range of different frameworks, and we have shown that in low silica frameworks there is a preference for the formation of discrete aluminium clusters. However, this is not the case for sodium-containing zeolites, where the global minimum structures are Löwensteinian ordered frameworks. In low silica sodium frameworks, next-nearest neighbour aluminium distribution is favoured, but next-next nearest neighbour distributions are preferred with increasing aluminium content. Marked differences between the most thermodynamically stable aluminium distributions of protonated and sodium-containing zeolites have been discerned demonstrating the influence of counter-cation identity on framework aluminium location. In addition, Dempsey's rule[30] is violated in the global minimum structures of all investigated frameworks.

The literature contains several reports of violations of Dempsey's rule in zeolites,[12,15,31] and it has been established that non-covalent interactions, present between framework oxygen and extra-framework cations, may distort aluminium distributions away from true Dempsey-ordering.[31] On close inspection, violations of Dempsey's rule in Na-SSZ-13 can be rationalised by simple electrostatics. As shown in the global minimum structure for 2 Al per unit cell (Figure 3a), there is a preference for $Na^+$ cations to maximise their coordination with framework oxygen whilst minimising unfavourable cation-cation interactions – as illustrated by the collection of unusually high energy structures (with aluminium separations of 5·80 - 8·20 Å) in Figure 2a, all of which contain Na$^+$ cations at relatively unfavourable low separations, causing these structures to be destabilised compared with what would be expected from Dempsey's rule. The importance of Na$^+$ cations in determining the distribution of framework aluminium throughout a zeolite is also reflected in the variation in the position of the third aluminium for the two Na-SSZ-13 structures with 3 Al per unit cell. In the Löwensteinian structure, the NNN Al position is favoured, and the associated Na$^+$ cations occupy two 8-rings and one 6-ring, with minimal repulsions between the counter-cations. However, the NNNN Al position is favoured for the non-Löwensteinian structure, in which the Na$^+$ cations occupy only one 8-ring, and the two 6-rings of the D6R. In this structure, a single 6-ring and 8-ring occupancy are filled by virtue of the initial non-Löwensteinian arrangement of the aluminium ions. NNN substitution would result in Na$^+$ occupancy of a six-ring that is already filled. Despite the NNNN position traditionally being thought of as more unfavourable, in the NL case, it is the only aluminium position which can satisfy the Na$^+$ coordination requirements whilst minimising unfavourable Na-Na interactions.

Rationalising the non-Dempsey aluminium distribution in sodium-containing frameworks is straightforward, whilst untangling the thermodynamic preference for NL ordering in protonated frameworks is more complex. As demonstrated by sodium-containing frameworks, non-covalent interactions play a significant role in determining aluminium distribution, we hence speculated that hydrogen-bonding interactions could be the cause of the unanticipated stability of the NL ordering in protonated zeolite frameworks. Fujita *et al.* demonstrated that hydrogen bonding interactions cause aluminium atoms to reside in close proximity to one another in zeolite Beta.[31] The separation between framework oxygen and H1 and H2 in 2 Al per unit cell H-SSZ-13 indicates the existence of two hydrogen bonds (O-H---O < 2·5 Å) per aluminium in the both the global minimum structure and the lowest energy Löwensteinian structure. We hence examined the robustness of the order of stability predicted in this work by using other density functionals. A representative subset of structures were selected and re-optimised with the revPBE[32] and BLYP functionals,[33,34] which have been shown to underbind hydrogen bonding interactions in water and ice structures (whilst PBE overbinds).[35] The results (S1) show that decreasing the hydrogen bonding strength in this way has no qualitative effect on the results and little quantitative effect, indicating that whilst hydrogen bonding must play a part in stabilising the H-SSZ-13 structures, it is not the decisive factor that controls whether NL is favoured over L.

Next we considered the charge distributions in the structures. On comparison of the Mulliken population analysis for 2 Al per unit cell Na and H-SSZ-13 we found the charge on the Na$^+$ cation is +1·07 (Hirshfeld = +0.97), far greater than that of the proton in corresponding H-SSZ-13 NL structure H$^+$ = +0·42 (Hirshfeld = +0·27). Consequently, the charge on the framework oxygen atoms covalently bound to the protons is substantially reduced in comparison to the corresponding oxygen atoms in the Na-SSZ-13 structure, O = -0·71 versus -1·08 (Hirshfeld = -0·68 and -0·79) in the H-SSZ-13 and Na-SSZ-13 structures, respectively. The reduced charge on the oxygen in the protonated case, results in the formation of long T-O(H) bonds (where T is Si or Al). According to the DFT data, the Al-O bond is more deformable than Si-O, in line with expectation as the absolute charge on Al is smaller than on Si and the radius of $Al^{3+}$ is greater than $Si^{4+}$. Hence, $Si^{4+}$ forms shorter, stronger more ionic bonds in comparison to $Al^{3+}$. For the H-SSZ-13 structure at 2 Al per U.C. Al-O(H) bonds are 11% longer compared to Al-O, stretched to a maximum of 1·90 Å, compared to an average bond length of approximately 1·71 Å. Si-O(H) bonds are only 7% longer than Si-O, stretched to a maximum bond length of 1·74 Å, compared to a framework average of approximately 1·63 Å. Each of the stable NL H-SSZ-13 structures contain a hydroxyl species mediating two aluminium ions, these structures therefore contain a total of three long Al-O(H) bonds, and a single long silanol Si-O(H) bond. In the high energy NL structures and all L structures, there are two Al-O(H) bonds, and two comparatively unfavourable long Si-O(H) bonds. Adopting the NL configuration minimises the the number of long Si-O(H) bonds and maximises the number of the short, strong, more ionic Si-O bonds. In the sodium loaded zeolites, the interaction between $Na^+$ and framework oxygen is primarily electrostatic and there is essentially complete charge transfer between Na and framework oxygen, as reflected by the computed Na charge and so the difference in ionicity/charge between a framework oxygen coordinated to $Na^+$ and those not coordinated to to $Na^+$ is rather small. In H-SSZ-13, the electrons are smeared across the covalent O-H bond and the effective charge on the bridging oxygen is reduced and the alumina units favour adopting next-next-nearest neighbour structures. The clustering or islanding of aluminium has been noted in silicon-aluminium phosphate zeolites[12] but not in aluminosilicate zeolites.

To check whether the qualitative result is sensitive to the extra-framework cation, we performed further calculations, substituting $Na^+$ and $H^+$ cations in the 2 Al per unit cell SSZ-13 model, for intermediate sized $Li^+$ cations and optimised all configurations to equilibrium density. The DFT results are included in the SI (S3 and S4), and is remarkably similar to that of Na-SSZ-13, showing a thermodynamic preference for 'traditional' Löwensteinian ordering over non-Löwensteinian with a ΔE (NL$_{global\ minimum}$ – L$_{global\ minimum}$) = +51·2 kJmol$^{-1}$ per U.C. However, for Li-SSZ-13, the global minimum structure shows marked differences in alkali cation position, with Li$^+$ ions capping the faces of individual D6R units, rather than located at the parameter of the 8-rings, as was the case for Na-SSZ-13 at this Si/Al ratio. Because Li$^+$ cations are considerably smaller than Na$^+$ cations, the Li$^+$ ions are able to get closer to the D6R due to their higher charge density.

SSZ-13 is typically synthesised from a sodium solution with a nitrogenous structure-directing agent, yielding Na-SSZ-13, which is subsequently ion-exchanged post synthesis to give the protonated, Brønsted acid active form of the zeolite catalyst, H-SSZ-13.[36] It is this form of the zeolites that is used to catalyse methanol-to-olefin conversion, a proposed lucrative, non-petroleum route for the production of short-chain organic compounds. At present, there is no viable way to synthesise H-SSZ-13 directly, most probably due to the role of counter-cations in directing the progress of zeolite formation during synthesis. As shown by our results, the location of framework aluminium is directly affected by counter-cation identity, and we can therefore assume that the distribution of aluminium in the global minimum Na-SSZ-13 framework is most representative of what would likely be seen in typical samples of SSZ-13, as it is this cation which determines the position of Al. However, our predictions suggest that a direct synthesis of H-SSZ-13, H-LTA, H-RHO and H-ABW should favour NL aluminium ion ordering. Interestingly, high resolution mass spectrometry data concerning the incorporation of aluminium in prenucleating silicate species by Schaack et al.,[37] indicates that Löwenstein's rule is not obeyed for all silicate species. The work provides evidence of 4-ring units containing -Al-O-Al-, but concludes that whilst these species may occur in solution, species that obey Löwenstein's rule are preferentially formed. Nevertheless, the observation of pre-nucleating building units with -Al-O-Al- linkages hints that this motif may not be as elusive as is generally believed, and these sequences may be found in crystals.

Because direct synthesis of proton compensated zeolites has not yet been achieved, direct validation of NL ordered frameworks in protonated zeolites cannot be assessed immediately. However, with regard to the synthesis of H-zeolite frameworks, we propose that the formation of -Al-O-Al- might be facilitated *via* two post-synthetic methods. The first is to use water, which has been shown to facilitate the making and breaking of -Si-O-Si- and -Al-O-Si-.[38] Long term steeping of H-SSZ-13 in water could be expected to lead to the redistribution of Al in the framework, yielding -Al-O-Al- as the thermodynamically preferred arrangement. Potentially, very slightly acidic or basic water might enhance the rate of rearrangement without dealumination or desilication of the zeolite framework. A second potential approach is, in essence, reverse-dealumination; placing a zeolite crystal in a solution containing an excess of alumina units with the assumption that for high alumina zeolites, the aluminium content will rise, increasing the likelihood of alumina units situated adjacent to one another. Previously, this has been achieved in high-silica ZSM-5 *via* $AlCl_3$ vapour treatment, and in very low-silica zeolite Y using non-crystallisation inducing alkaline solutions (*e.g.* KOH) in the presence of large concentrations of extra-framework aluminium. [39,40]

An intriguing question is whether the NL linkages that we have predicted are present in existing samples, and if so, what signatures could be used to unambiguously identify these -Al-O-Al- sequences. Recent atom tomography work[4,21] has vividly demonstrated that the distribution of aluminium in a typical ZSM-5 zeolite sample is very heterogeneous. At present there is no available method that can accurately distinguish framework aluminium from framework silicon with Ångström resolution.[4] A 2010 work by Wright *et al.*[8] concerning possible non-Löwensteinian structures observed in gallosilicates, discusses the possibility of using of $^{17}$O magic angle spinning (MAS) NMR to detect non-Löwensteinian ordering, a method which has been successful in identifying -Al-O-Al- linkages in aluminosilicate glasses.[41] We have examined the global minimum H-SSZ-13 structures (Si/Al = 17) and predicted $^{29}$Si, $^{27}$Al solid-state, MAS NMR shifts and IR frequencies, in an attempt to discern whether spectroscopic signatures exist that would be indicative of the presence of non-Löwensteinian ordering (see SI). However, at a Si/Al ratio of 17, typical for SSZ-13, the predicted $^{29}$Si NMR data shows that there is a slight decrease in the negativity of the chemical shift values for -Al-O-Al- containing frameworks. However, these shifts are well within the anticipated range for a zeolite at this Si/Al ratio, and far too similar to the chemical shifts of the surrounding Si atoms to be used practically as a characterisation method. Similarly, predicted vibrational frequencies indicate that characteristic stretches would not be detectable due to overlap of Al-O(H)-Al stretches with that of Si-O(H)-Si and Si-O(H)-Al, and -Al-O-Al- stretches with Si-O-Al. This data is included and discussed in the SI (S8 and S9).

If the proposed post-synthetic techniques or alternative synthetic strategies are successful in realising zeolites with NL framework aluminium distributions, as predicted by this work, these materials would be potentially invaluable for the development of new zeolite catalysts. Despite the advantages of using zeolites in catalysis, for example, specificity and size exclusion properties, it is well documented that the catalytic efficiency of microporous materials is often limited by restricted access to active sites. Introducing ordered, controllable meso- and macroporosity to the framework provides a solution to mass transport limited diffusion through the porous zeolite network, the introduction of hierarchy has also been shown not only enhance catalytic activity, but also stability in a range of zeolite frameworks. A variety of of both bottom-up and top-down strategies have proved successful for hierarchically ordered zeolite synthesis. The post-synthetic introduction of mesoporosity by the extraction of framework atoms is a particularly popular method, and can be achieved by acid, base or steam treatment of the zeolite material.[42,43] One can imagine how techniques such as these could be used to dealuminate low-silica aluminium cluster-containing materials, similar to those predicted in this work. For example, removing all four alumina units in the 4 Al per unit cell H-SSZ-13 global minimum structure predicted by DFT would increase the 7 Å, 8-ring aperture cavity system, with a void-space of approximately 10 Å in diameter, to up to 17 Å, approaching mesoporosity. Crucially, the calculations indicate that not only is aluminium clustered, but it is also located in predictable, ordered positions, which suggests that introduced porosity *via* selective dealumination could be controllable in H-zeolites.

The reaction mechanisms and deactivation pathways of real catalytic zeolite materials is relatively a poorly understood area of zeolite science, not withstanding remarkable recent advances.[44,45] In part this is due to a lack of molecular-level information concerning the location of framework alumina and associated counter-cations, which are thought to be integral to the catalytic reaction mechanism. Clustering of aluminium and the associated clustered acid sites, as predicted by the DFT results, is suggestive of new, potentially more reactive sites (due to the density of acid sites) and new reaction pathways which have not yet been considered. Perea *et al.* have shown that the Si/Al distribution can be inhomogeneously distributed throughout the zeolite framework,[4] so it is conceivable that aluminium cluster motifs, including non-Löwensteinian linkages already exist in real zeolite materials, and may impact reaction and deactivation pathways operating in current catalysts.

The realisation of zeolite materials with contradistinct aluminium distributions to those synthesised by traditional routes holds enormous potential for the future of zeolite catalysis. We hope this work stimulates experimental investigation into the direct or post-synthesis of non-Löwensteinian ordered zeolites and further characterisation of existing materials.

**Methods**

The majority of the periodic DFT calculations were performed using the CP2K code,[23–25] and additional benchmark calculations for energetics and solid-state NMR were performed using the CASTEP code.[46] Results were calculated using the PBE[26] functional, although further calculations using revPBE[32] and BLYP[33,34] were included to verify our initial 2 Al per unit cell SSZ-13 findings. These calculations, and methods, are discussed in detail in the supplementary information, along with a full description of the single-point energy PBE0[28,29] calculations mentioned in our results and discussion. All framework structures were obtained in their all-silica form from the database of zeolite structures,[22] and permutations of 2, 3 and 4 Al per unit cell models created by the methodology discussed in the main body of the text. Individual models were fully geometry optimized to equilibrium density, with variable lattice parameters in CP2K as 1:1:1 cells using the high quality TZV2P basis set and an energy cutoff of 650 Ry. Only the ABW framework was optimized as a 2:2:2 supercell, due to its small unit cell size. We also tested a selection of larger 2:2:2 supercells for each of the frameworks, although we saw no meaningful change in the relative energies using the larger cells. Additional computational details are included in the Supporting Information.

**Acknowledgments**

We thank Dr David McKay for helpful discussions on solid-state NMR calculations. This work was made possible through the UK's HEC Materials Chemistry Consortium, which is funded by EPSRC (EP/L000202), and the use of the ARCHER UK National Supercomputing Service (http://www.archer.ac.uk).

**References**

1. Vermeiren, W. & Gilson, J-P. *Top. Catal.* **52,** 1131–1161 (2009).

2. Guisnet, M. & Gilson, J-P. Zeolites for Cleaner Technologies, (Imperial College Press, 2002).

3. Anastas, P. T. & Warner, J. C. Green Chemistry: Theory and Practice, (Oxford University Press, 1998).

4.  Perea, D. E. *et al. Nat. Commun.* **6,** 7589 (2015).

5.  Loewenstein, W. *Am. Mineral.* **39,** 92–96 (1954).

6.  Bursill, L. A., Lodge, E. A., Thomas, J. M. & Cheetham, A. K. *J. Phys. Chem* **85,** 2409–2421 (1981).

7.  Klinowski, J., Thomas, J. M., Fyfe, C. A. & Hartman, J. S. *J. Phys. Chem* **85,** 2590–2594 (1981).

8.  Shin, J. *et al. Dalt. Trans.* **39,** 2246 (2010).

9.  Tarling, S. E., Barnes, P. & Klinowski, J. *Acta Cryst.* **37,** 652–636 (1980).

10. Bell, R. G., Jackson, R. A. & Catlow, C. R. A. *Zeolites* **70,** 870–871 (1992).

11. Catlow, C. R. A., George, A. R. & Freeman, C. M. *Chem. Commun.* 1311–1312 (1996).

12. Akporiaye, D. E., Dahl, I. M., Mostad, H. B. & Wendelbo, R. *J. Phys. Chem.* **100,** 4148–4153 (1996).

13. Melchior, M., Vaughan, D. & Jacobson, A. *J. Am. Chem. Soc.* **104,** 4859–4864 (1982).

14. Soukoulist, C. M. *J. Phys. Chem* 4898–4901 (1984).

15. Schroeder, K. P. & Sauer, J. *J. Phys. Chem.* **97,** 6579–6581 (1993).

16. Yang, C.-S., Mora-Fonz, J. M. & Catlow, C. R. A. *J. Phys. Chem. C* **115,** 24102–24114 (2011).

17. Lo, C. & Trout, B. *J. Catal.* **227,** 77–89 (2004).

18. Muraoka, K., Chaikittisilp, W. & Okubo, T. *J. Am. Chem. Soc.* **138,** 6184–6193 (2016).

19. Zokaie, M., Olsbye, U., Lillerud, K. P. & Swang, O. *Microporous Mesoporous Mater.* **158,** 175–179 (2012).

20. Gábová, V. *et al. Chem. Commun.* **11,** 1196–1197 (2003).

21. Schmidt, J. E. *et al. Angew. Chemie - Int. Ed.* **55**, 11173–11177 (2016).

22. Baerlocher, C., McCusker, L. & Olson, D. H., Atlas of zeolite framework types (6th Edition), Elsevier, (2007).

23. Available at: https://www.cp2k.org/.

24. VandeVondele, J. *et al. Comput. Phys. Commun.* **167,** 103–128 (2005).

25. VandeVondele, J. & Hutter, J. *J. Chem. Phys.* **127,** 114105 (2007).

26. Perdew, J. P., Burke, K. & Ernzerhof, M. *Phys. Rev. Lett.* **78,** 1396–1396 (1997).

27. Dempsey, E., Kuehl, G. H. & Olson, D. H. *J. Phys. Chem.* **73,** 387–390 (1969).

28. Adamo, C. & Barone, V. *J. Chem. Phys.* **110,** 6158 (1999).

29. Guidon, M., Hutter, J. & VandeVondele, J. *J. Chem. Theory Comput.* **6,** 2348–2364 (2010).

30. Dempsey, E. *J. Catal.* **1**, 115–119 (1977).

31. Fujita, H., Kanougi, T. & Atoguchi, T. *Appl. Catal. A Gen.* **313,** 160–166 (2006).

32. Zhang, Y. & Yang, W. *Phys. Rev. Lett.* **80,** 890–890 (1998).

33. Becke, A. D. *Phys. Rev. A* **38,** 3098–3100 (1988).

34. Lee, C., Yang, W. & Parr, R. G. *Phys. Rev. B* **37,** 785–789 (1988).

35. Gillan, M. J., Alfè, D. & Michaelides, A. *J. Chem. Phys.* **144,** 130901 (2016).

36. Zones, S. I. US Patent 4544538 (1985).

37. Schaack, B. B. PhD Thesis (Ruhr Universität Bochum, 2009).

38. Von Ballmoos, R. & Meier, W. M. *J. Phys. Chem.* **86,** 2698–2700 (1982).

39. Liu, X., Klinowski, J. & Thomas, J. M. *J. Chem. Soc., Chem. Commun.,* 582-584 (1986).

40. Bezman, R. D. *J. Chem. Soc., Chem. Commun.* 1562–1563 (1987).

41. Stebbins, J. F., Lee, S. K. & Oglesby, J. V. *Am. Mineral.* **84**, 983–986 (1999).

42. Verboekend, D. & Pérez-Ramírez, J. *Catal. Sci. Technol.* **1**, 879–890 (2011).

43. Möller, K. & Bein, T. *Science (80),* **333**, 297 (2011).

44. de Smit, E. *et al. Nature* **456**, 222–225 (2008).

45. Buurmans, I. L. C. & Weckhuysen, B. M. *Nat. Chem.* **4,** 873–886 (2012).

46. Clark, S. J. *et al. Zeitschrift fuer Krist.* **220,** 567–570 (2005).

**Supporting Information: Violations of Löwenstein's rule in zeolites**

Rachel E. Fletcher, Sanliang Ling and Ben Slater*

Department of Chemistry, University College London, 20 Gordon Street, London

WC1H 0AJ, UK

*Email: b.slater@ucl.ac.uk

**Index**

**S1** The relative of stabilities of a collection of high-, mid-, and low-energy 2 Al per unit cell (Si/Al = 17) H-SSZ-13 (CHA) configurations, geometry optimized using CP2K with PBE, BLYP and revPBE functionals, and CASTEP at the PBE level of theory.[1–5]

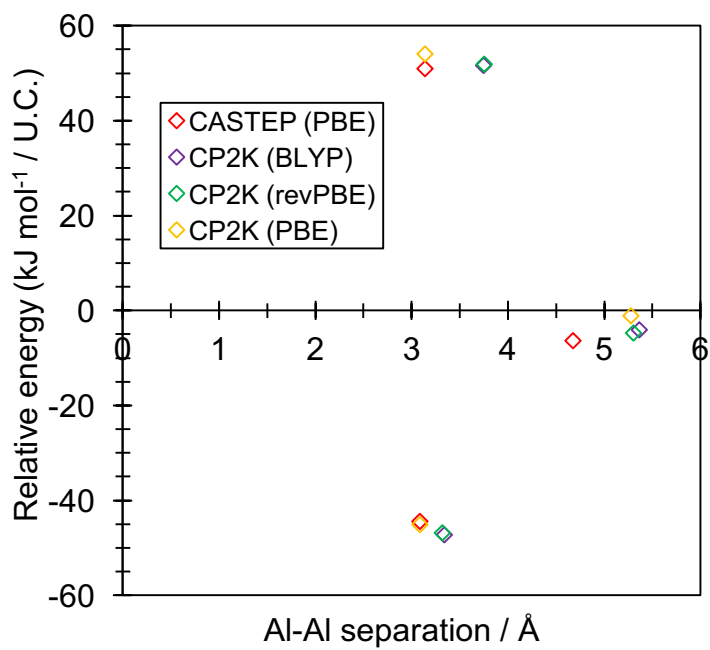

Energy differences between the three frameworks appear to be consistent. Discrepancies in Al-Al separation for the same models between the different levels of theory is due to differences in framework density following geometry optimization.

**S2** Relative stabilities of the 12 most stable 2 Al per unit cell (Si/Al = 17) H-SSZ-13 (CHA) structures, including both NL and L configurations, calculated using CP2K at the PBE level of theory and hybrid functional PBE0, implemented in CP2K using auxiliary density matrix methods (ADMM) [6,7]

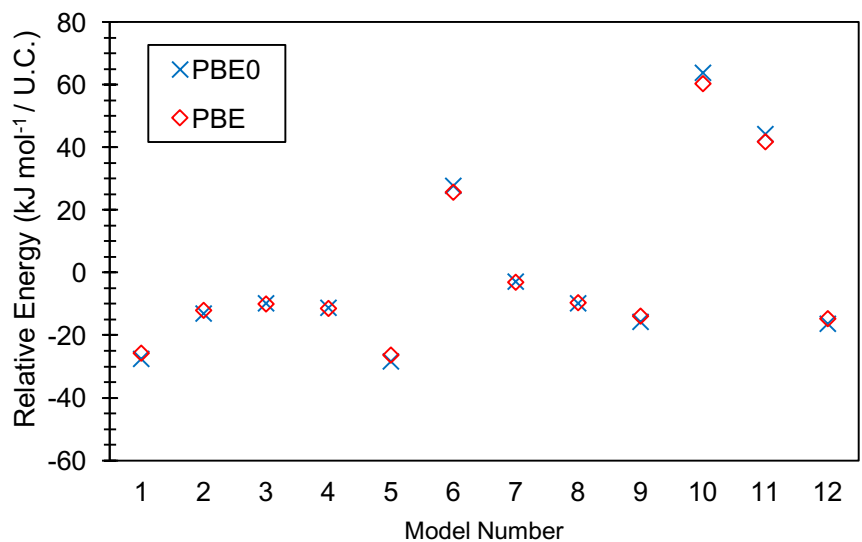

The PBE0 data points are single point energies performed on the PBE relaxed configurations. There is good correlation between data obtained using the PBE functional, and the more sophisticated, more computationally demanding, hybrid PBE0 functional, such that many of the points shown on the graph above are in fact coincidental. Clearly, the PBE data captures the qualitative and quantitative differences between different configurations.

**S3 Relative energy dispersion (kJ mol⁻¹ per U.C.) against framework aluminium separation for Li-SSZ-13 Frameworks possessing non-Löwensteinian ordered aluminium atoms (-Al-O-Al-) are shown in blue.**

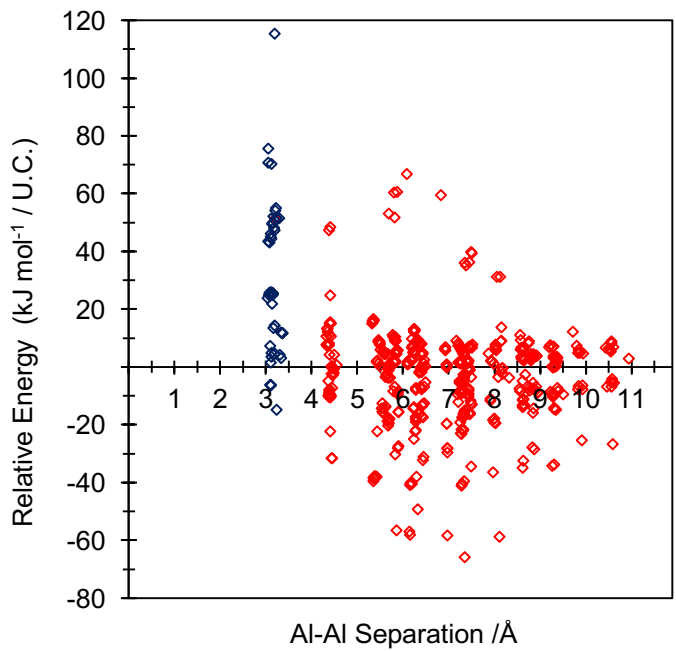

**S4 Global minimum structure according to DFT for Li-SSZ-13 with 2 Al per unit cell (Si/Al = 17). A Löwensteinian structure with Li⁺ cations above and below the planes of the six-rings. Si are shown in yellow, O in red, Al in light blue and Li in dark blue.**

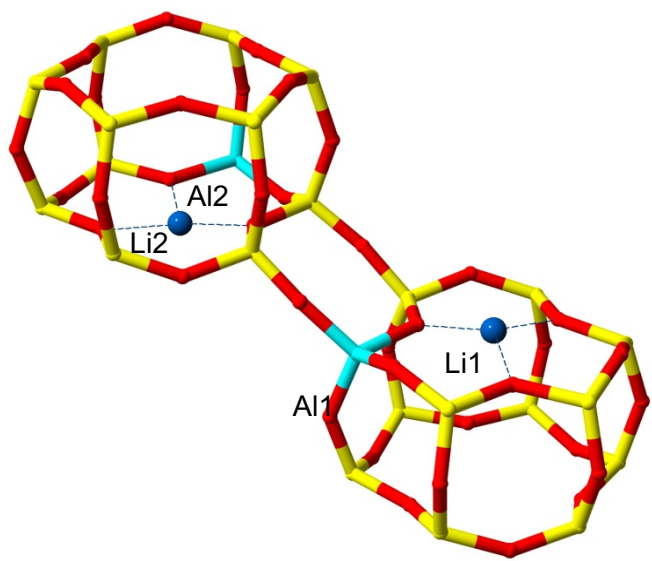

**S5** Löwensteinian global minimum structures according to DFT for Na-SSZ-13 with 3 Al per unit cell (Si/Al = 33) originating from the H-SSZ-13 2 Al per unit cell NL and L global minimum structures, and 4 Al per unit cell (Si/Al = 32) originating from the H-SSZ-13 NL global minimum structure.
A) NL 3Al per unit cell Na-SSZ-13    B) L 3 Al per unit cell Na-SSZ-13    C) 4 Al per unit cell Na-SSZ-13

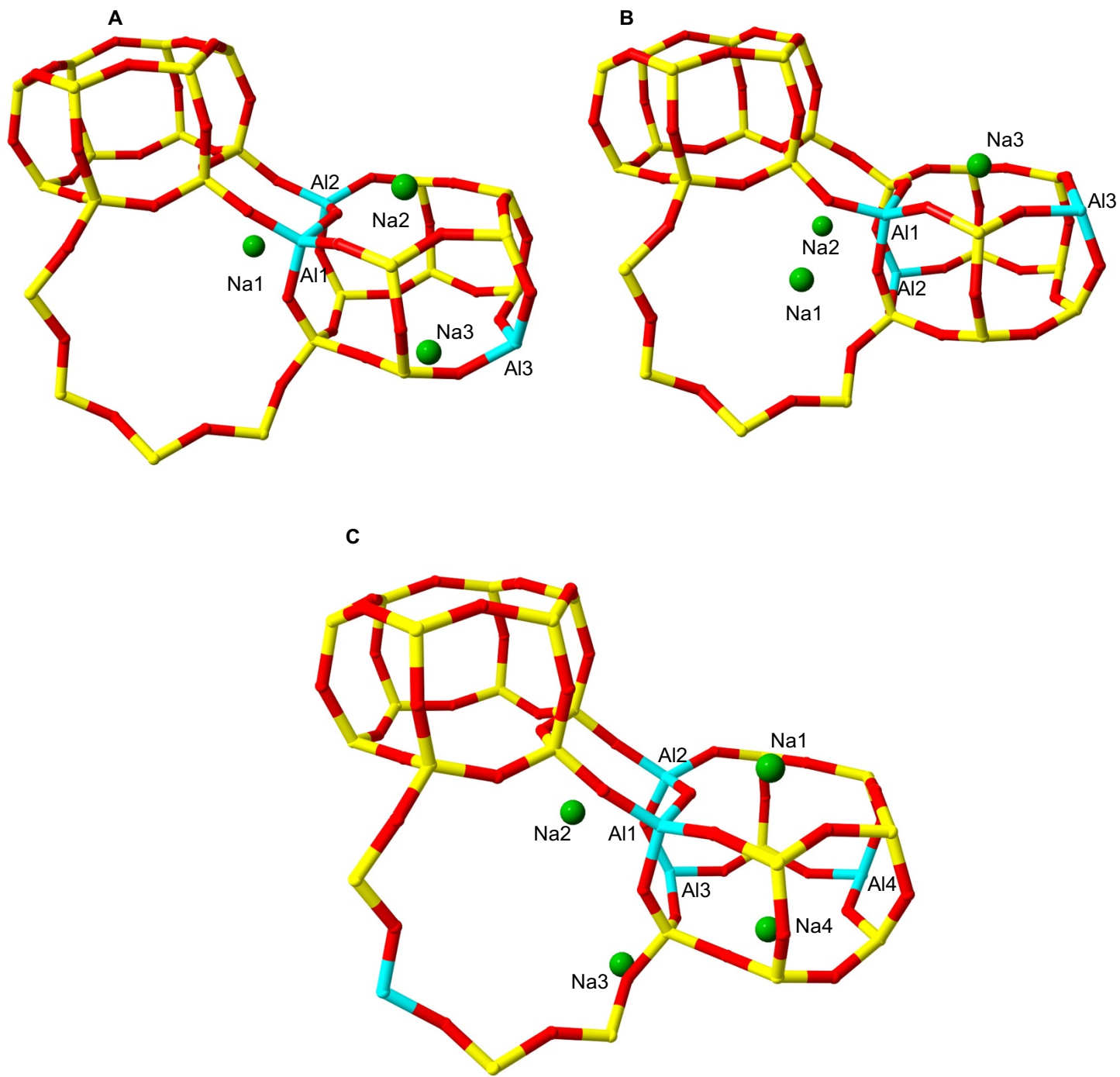

**S6 Framework topologies included in this study; A) LTA, B) RHO and C) ABW, and table of their corresponding framework densities and composite building units.** [8]

| Framework | Density / Tnm⁻¹ | Composite building units |
|---|---|---|
| CHA | 15.1 | *d6r, cha-cage* |
| LTA | 14.2 | *d4r, sodalite-cage, lta-cage* |
| RHO | 14.5 | *d8r, lta-cage* |
| ABW | 17.6 | *abw-cage* |

**A**

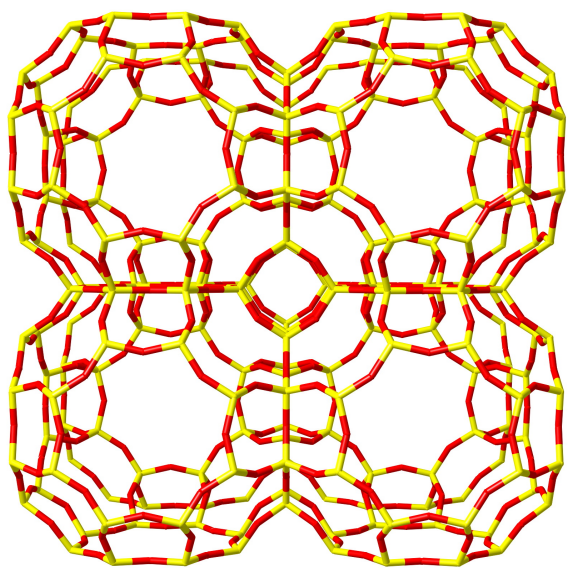

**B**

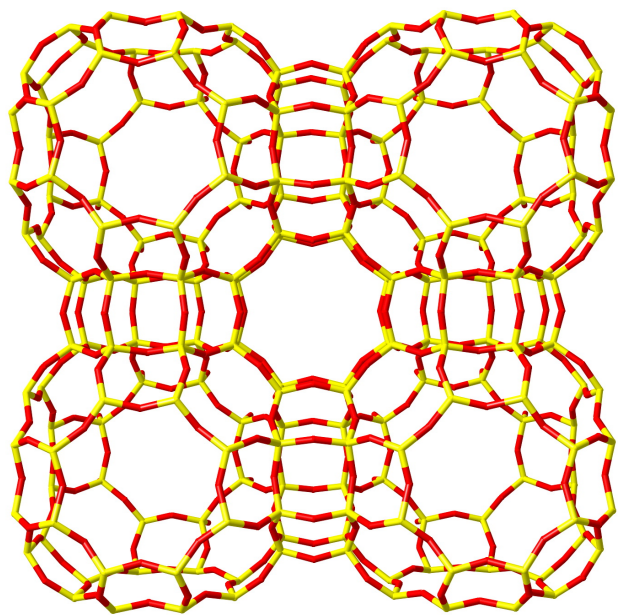

**C**

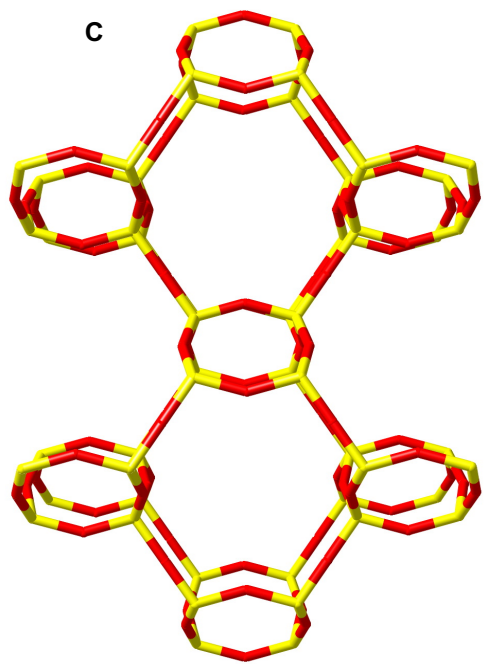

**S7 Non-Löwensteinian global minimum structures according to DFT for RHO, LTA, ABW-type frameworks in their protonated forms with 2 Al per unit cell.**
**A) LTA B) RHO C) ABW**

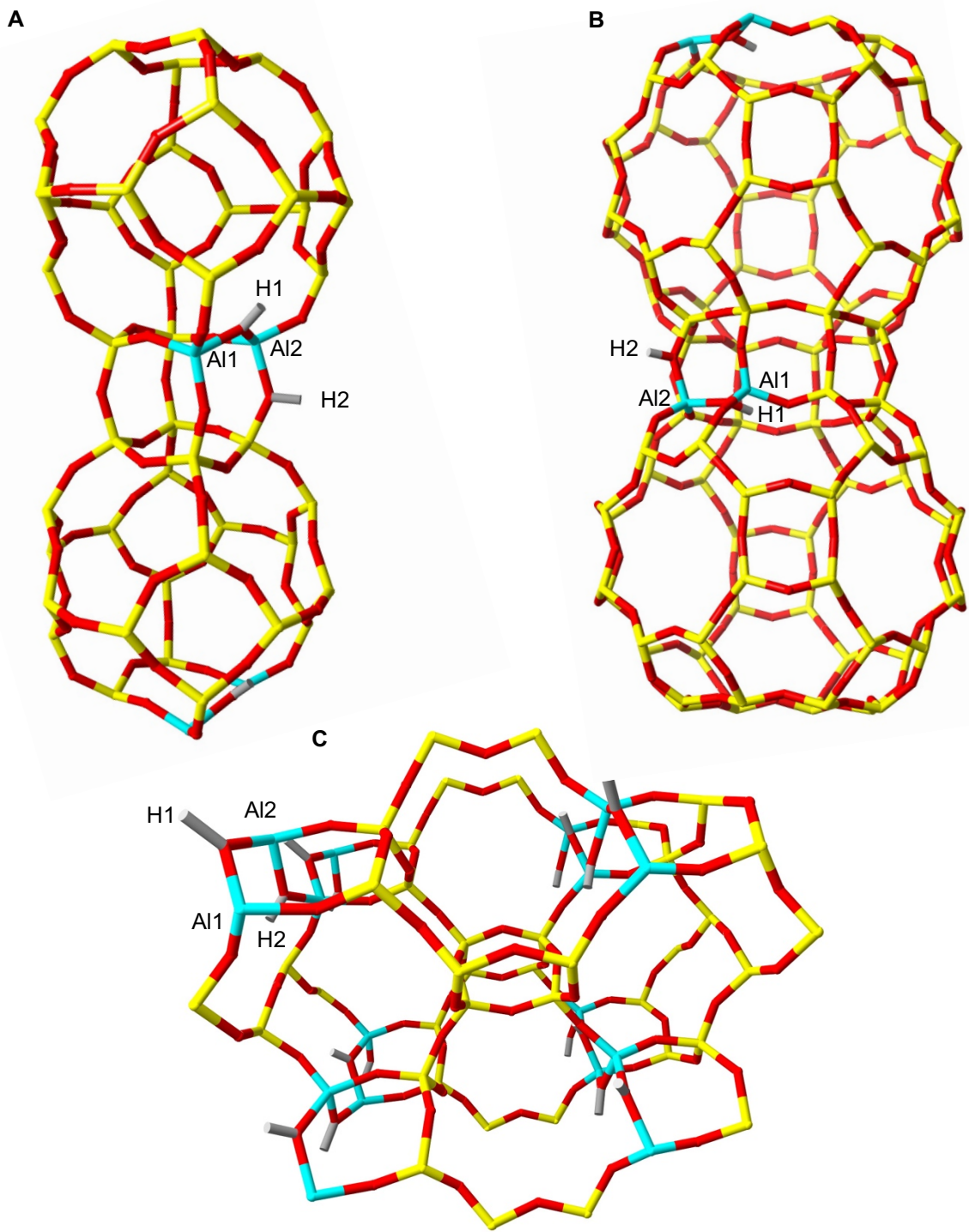

# S8 SS MAS <sup>29</sup>Si NMR Data

Solid-state NMR calculations were performed for the L and NL 2 Al per unit cell global minimum structures using CASTEP (version 8.0)[5] with the CP2K optimized geometry because of the high computational cost of geometry optimisations of large systems associated with planewave basis sets. Our tests on a similar porous system, the UiO-66 metal-organic framework, showed that the difference in calculated NMR chemical shifts due to minor geometry differences between different DFT functionals and computational codes can indeed be safely neglected. All CASTEP calculations were performed using the PBE functional, on-the-fly pseudopotentials and planewave basis sets with a cutoff of 60 Ry, and a Monkhorst-Pack *k*-points grids of (3x3x3) were used to sample the Brillouin zone. The $^{29}$Si chemical shifts, referenced to tetramethylsilane (TMS), are shown below.

The anticipated chemical shifts for different silica environments are as follows:[9]

| 4Si (0Al) | 3Si (1Al) | 2Si (2Al) | 1Si (3Al) | 0Si (4Al) |
|---|---|---|---|---|
| -100 to -115 ppm | -96 to -107 ppm | -91 to -100 ppm | -85 to -95 ppm | -80 to -91 ppm |

Increasing the amount of aluminium bonded to the silica tetrahedra significantly decreases the negativity of the chemical shift (shifted downfield). This phenomenon is observed in our DFT predictions. The average chemical shift of 4Si in both frameworks is approximately -114 ppm, this increases to a maximum shift of -104.9 ppm for SiO$_4$ bonded to a single aluminium in the NL structure. Our predictions show that this method could not be used to characterize -Al-O-Al- at such a high Si/Al ratio. Firstly, the chemical shifts for all silica environments are far too similar, also any peaks that could be considered 'characteristic' of a nearby –Al-O(H)-Al- would be lost in background noise in the NMR spectrum.

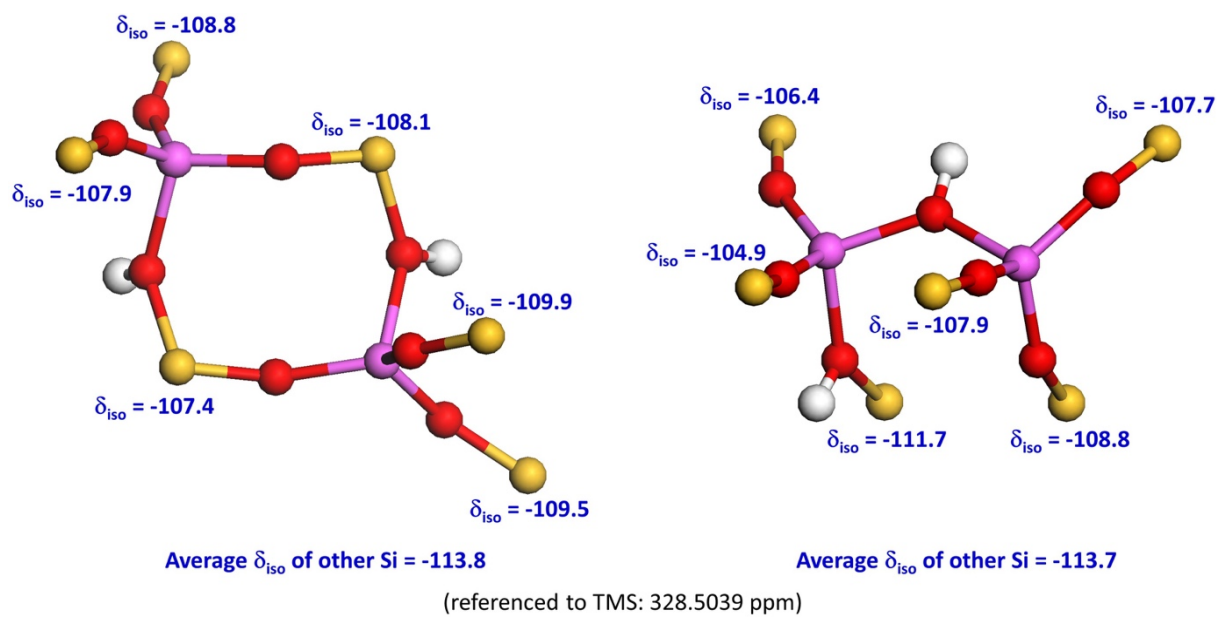

(referenced to TMS: 328.5039 ppm)

## S9 Vibrational frequencies

Using CP2K, we predicted the vibrational frequencies of the L and NL 2 Al / unit cell H-SSZ-13 global minimum structures. The vibrational frequencies are shown below, the stretches indicative of a free hydroxyl bonded at –Si-O-Al- and -Al-O-Al- (3500 - 3700 cm$^{-1}$) are highlighted in grey. These stretches overlap with other broad stretches and would not be uniquely discernible.

Vibrational Frequencies / cm$^{-1}$

| NL H-SSZ-13 | | | | L H-SSZ-13 | | | |
|---|---|---|---|---|---|---|---|
| 41.086 | 280.913 | 459.911 | 775.480 | 34.079 | 276.316 | 451.709 | 773.324 |
| 60.700 | 284.215 | 464.448 | 776.906 | 55.845 | 279.429 | 453.624 | 774.907 |
| 62.841 | 285.040 | 466.106 | 780.052 | 58.215 | 281.211 | 457.809 | 778.264 |
| 66.204 | 286.070 | 467.836 | 780.401 | 67.772 | 282.933 | 462.642 | 779.523 |
| 71.213 | 288.480 | 470.124 | 794.045 | 69.986 | 285.094 | 464.906 | 793.011 |
| 79.036 | 291.250 | 470.484 | 796.538 | 73.962 | 286.222 | 467.940 | 794.640 |
| 83.716 | 293.592 | 472.992 | 800.852 | 77.216 | 288.777 | 469.025 | 798.655 |
| 91.913 | 295.027 | 476.505 | 974.877 | 79.369 | 289.208 | 470.884 | 993.773 |
| 93.805 | 296.247 | 477.041 | 977.147 | 84.305 | 290.988 | 473.611 | 1002.309 |
| 97.372 | 297.325 | 479.247 | 998.464 | 91.581 | 293.718 | 476.569 | 1003.629 |
| 101.567 | 298.764 | 480.501 | 1001.794 | 95.905 | 296.944 | 478.007 | 1008.455 |
| 103.036 | 301.419 | 480.755 | 1005.069 | 99.897 | 297.771 | 480.059 | 1009.491 |
| 109.587 | 302.791 | 482.842 | 1008.935 | 104.715 | 299.632 | 481.547 | 1012.134 |
| 112.457 | 303.172 | 484.444 | 1013.073 | 107.956 | 302.274 | 483.072 | 1014.181 |
| 115.894 | 305.490 | 487.453 | 1015.120 | 109.689 | 303.286 | 486.479 | 1015.945 |
| 122.126 | 308.469 | 490.901 | 1017.229 | 114.726 | 306.703 | 487.946 | 1017.846 |
| 123.307 | 310.304 | 491.421 | 1019.178 | 117.631 | 306.916 | 490.320 | 1018.454 |
| 127.511 | 314.905 | 493.762 | 1020.337 | 120.582 | 309.430 | 493.185 | 1019.376 |
| 134.077 | 317.646 | 495.863 | 1021.235 | 128.715 | 310.100 | 497.700 | 1022.931 |
| 134.407 | 319.222 | 499.681 | 1022.248 | 131.484 | 314.976 | 501.279 | 1024.867 |
| 137.322 | 320.221 | 506.260 | 1024.148 | 134.693 | 316.073 | 505.999 | 1027.613 |
| 142.557 | 324.034 | 509.776 | 1024.340 | 137.829 | 318.730 | 507.753 | 1028.312 |
| 145.870 | 326.486 | 511.175 | 1027.749 | 138.905 | 319.934 | 509.569 | 1029.497 |
| 147.035 | 330.468 | 515.541 | 1029.806 | 142.454 | 326.546 | 511.889 | 1031.092 |
| 147.733 | 331.858 | 536.236 | 1030.197 | 144.373 | 328.034 | 528.920 | 1032.354 |
| 151.114 | 333.907 | 544.018 | 1031.939 | 144.928 | 329.369 | 531.849 | 1033.877 |
| 152.361 | 337.729 | 551.341 | 1032.664 | 150.270 | 331.673 | 546.898 | 1034.714 |
| 156.749 | 338.623 | 563.546 | 1035.630 | 154.458 | 332.887 | 553.237 | 1036.601 |
| 158.904 | 340.575 | 568.779 | 1036.356 | 156.840 | 334.536 | 568.751 | 1039.284 |
| 160.488 | 341.975 | 583.620 | 1037.189 | 158.086 | 337.647 | 586.200 | 1039.398 |
| 161.486 | 342.371 | 591.713 | 1038.148 | 161.620 | 338.406 | 588.652 | 1040.573 |
| 165.783 | 344.140 | 593.259 | 1038.653 | 164.384 | 339.932 | 592.987 | 1041.621 |
| 166.708 | 347.491 | 596.488 | 1042.923 | 166.364 | 342.812 | 594.907 | 1044.496 |
| 170.959 | 350.620 | 598.476 | 1043.823 | 168.419 | 347.192 | 598.416 | 1045.304 |
| 172.948 | 352.842 | 601.131 | 1044.487 | 169.194 | 347.824 | 598.575 | 1047.497 |
| 174.223 | 354.989 | 601.522 | 1047.496 | 173.004 | 350.606 | 600.278 | 1047.930 |
| 177.640 | 358.478 | 608.647 | 1049.074 | 176.228 | 352.831 | 601.513 | 1051.638 |
| 178.923 | 360.613 | 612.718 | 1050.657 | 176.860 | 353.919 | 607.803 | 1053.262 |
| 182.899 | 361.603 | 614.399 | 1052.260 | 179.789 | 361.705 | 608.873 | 1054.064 |
| 183.027 | 363.761 | 616.432 | 1055.648 | 184.165 | 363.032 | 617.486 | 1055.753 |
| 184.783 | 367.420 | 620.433 | 1059.292 | 185.135 | 364.830 | 619.851 | 1056.985 |
| 185.955 | 371.459 | 624.156 | 1066.223 | 185.732 | 367.279 | 620.922 | 1073.847 |
| 188.620 | 373.127 | 627.769 | 1073.359 | 187.387 | 368.186 | 623.672 | 1082.761 |
| 190.671 | 374.184 | 637.823 | 1083.783 | 188.126 | 371.972 | 642.725 | 1087.536 |
| 192.691 | 376.490 | 652.838 | 1103.831 | 190.826 | 372.989 | 652.676 | 1099.736 |
| 193.738 | 377.708 | 666.989 | 1111.319 | 191.002 | 373.270 | 656.263 | 1110.110 |
| 194.571 | 379.022 | 677.882 | 1121.494 | 193.120 | 374.284 | 672.431 | 1116.757 |
| 195.994 | 380.609 | 680.600 | 1122.121 | 194.137 | 377.214 | 676.902 | 1117.905 |
| 196.682 | 382.328 | 685.354 | 1123.747 | 194.497 | 378.374 | 678.813 | 1121.236 |
| 198.895 | 383.619 | 686.535 | 1125.932 | 197.730 | 380.332 | 682.719 | 1126.825 |
| 200.963 | 385.723 | 693.316 | 1128.760 | 198.988 | 381.666 | 693.006 | 1128.819 |
| 202.850 | 386.332 | 697.145 | 1131.447 | 202.576 | 382.311 | 708.745 | 1129.863 |
| 203.654 | 387.263 | 713.096 | 1135.988 | 203.254 | 385.485 | 722.359 | 1134.511 |
| 204.135 | 388.917 | 730.237 | 1137.010 | 206.637 | 385.882 | 728.002 | 1135.050 |
| 207.644 | 391.012 | 734.399 | 1138.314 | 208.239 | 387.447 | 733.763 | 1140.596 |
| 209.803 | 391.441 | 738.851 | 1142.604 | 208.841 | 387.864 | 734.614 | 1142.541 |
| 213.123 | 394.794 | 742.477 | 1143.669 | 214.696 | 389.946 | 740.692 | 1146.420 |
| 216.380 | 395.470 | 746.615 | 1145.350 | 219.292 | 392.472 | 743.545 | 1147.640 |
| 221.024 | 397.282 | 748.056 | 1148.473 | 219.734 | 393.165 | 748.788 | 1148.864 |
| 224.110 | 399.088 | 750.353 | 1150.856 | 220.819 | 393.821 | 749.008 | 1152.122 |
| 226.453 | 401.381 | 751.393 | 1151.481 | 221.968 | 398.271 | 749.974 | 1152.560 |
| 228.693 | 402.226 | 752.215 | 1155.128 | 226.289 | 399.384 | 751.049 | 1156.436 |
| 230.187 | 403.995 | 752.989 | 1156.439 | 229.192 | 401.921 | 752.203 | 1158.169 |
| 232.893 | 405.128 | 754.186 | 1158.231 | 232.483 | 403.626 | 752.689 | 1159.746 |
| 235.102 | 407.200 | 754.979 | 1158.909 | 238.339 | 405.894 | 753.368 | 1163.000 |
| 245.470 | 413.514 | 756.073 | 1161.212 | 239.145 | 413.223 | 754.148 | 1164.140 |
| 247.639 | 413.935 | 756.288 | 1163.527 | 244.171 | 414.555 | 756.003 | 1165.520 |
| 250.655 | 416.097 | 757.296 | 1164.708 | 248.343 | 417.860 | 757.088 | 1165.981 |
| 252.351 | 418.391 | 759.423 | 1167.186 | 251.621 | 419.769 | 757.548 | 1167.974 |
| 258.163 | 421.317 | 760.947 | 1169.325 | 253.637 | 423.265 | 757.887 | 1170.740 |
| 261.616 | 422.868 | 763.969 | 1171.032 | 254.586 | 423.590 | 760.133 | 1172.795 |
| 263.343 | 425.553 | 765.144 | 1172.328 | 256.397 | 427.084 | 761.962 | 1174.619 |
| 264.785 | 428.809 | 766.028 | 1175.551 | 260.969 | 431.194 | 763.313 | 1176.196 |
| 266.483 | 432.927 | 766.922 | 1181.557 | 261.894 | 434.051 | 765.538 | 1177.260 |
| 268.482 | 436.148 | 767.268 | 1184.349 | 263.797 | 435.187 | 766.050 | 1179.158 |
| 270.691 | 437.957 | 767.892 | 1187.353 | 266.189 | 438.149 | 767.873 | 1181.042 |
| 272.142 | 438.603 | 769.125 | 1191.960 | 267.723 | 438.283 | 768.272 | 1185.312 |
| 273.041 | 438.751 | 769.933 | 1203.460 | 269.031 | 439.838 | 768.527 | 1190.400 |
| 275.187 | 442.348 | 770.510 | 1205.895 | 270.299 | 441.582 | 770.129 | 1201.193 |
| 276.693 | 445.236 | 772.746 | 3665.621 | 272.041 | 443.347 | 770.755 | 3696.139 |
| 276.938 | 448.423 | 773.513 | 3698.982 | 273.308 | 448.831 | 771.415 | 3699.446 |
| 277.645 | 454.582 | 774.920 | | 275.201 | 449.963 | 772.230 | |

**S10 Zero Point Energy Calculations and Thermodynamics Calculations**

Using the IR data (S7) we calculated the zero point energies for the 2 Al/ unit cell (Si/Al = 17) H-SSZ-13 L and NL global minima structures, obtained from DFT (CP2K) at the PBE level of theory, with the TZV2P basis set. Zero point energy = ZPE.

ZPE for NL = 1123.3 kJ mol$^{-1}$
ZPE for L = 1119.9 kJ mol$^{-1}$
$\Delta$ ZPE (NL-L) = 3.43 kJ mol$^{-1}$
$\Delta$E (NL-L) = -14.21 kJ mol$^{-1}$
$\Delta$E (NL-L) + ZPE = -10.78 kJ mol$^{-1}$

We then compared these results with data obtained using CASTEP for the smaller 2 Al per unit cell 12 T-site rhombohedral unit cell, (Si/Al = 5).

ZPE for NL = 398.4 kJ mol$^{-1}$
ZPE for L = 398.20 kJ mol$^{-1}$
$\Delta$ ZPE (NL-L) = 0.20 kJ mol$^{-1}$
$\Delta$E (NL-L) = -9.06 kJ mol$^{-1}$
$\Delta$E (NL-L) + ZPE = -8.86 kJ mol$^{-1}$

The absolute zero point energy difference between the CASTEP data with a 10 T atom unit cell are slightly lower than that obtained using CP2K for the larger 36 T atom unit cell. This is due to a combination of the lower Si/Al ratio and differences in the pseudopotential and basis sets used. The key point is that the enthalpy difference for NL-L still favours NL by ~10 kJ /mol when corrections for the ZPE are considered.

CASTEP was used to carry out an assessment of the free difference, taking account of the vibrational entropy differences within the static, harmonic approximation with a cutoff of 800 eV. Phonon frequencies were sampled at 4 different k-points in the Brillouin zone.

According to CASTEP, the internal energy difference between the Si/Al = 5 global minimum configurations is $\Delta$E(NL-L) = -9.06 kJ/mol. The Helmholtz free-energy difference between configurations as a function of temperature is reported below. The NL configuration is marginally stabilised with respect to L with increasing temperature.

| TEMPERATURE / K | HELMHOLTZ FREE ENERGY DIFFERENCE (NL-L) INCLUDING ZERO POINT ENERGY /kJmol$^{-1}$ |
|---|---|
| 0 | -8.86 |
| 37 | -8.89 |
| 73 | -8.99 |
| 110 | -9.10 |
| 147 | -9.21 |
| 183 | -9.32 |
| 220 | -9.43 |
| 257 | -9.53 |
| 293 | -9.63 |
| 330 | -9.73 |
| 367 | -9.83 |
| 403 | -9.93 |
| 440 | -10.03 |
| 476 | -10.12 |
| 513 | -10.22 |
| 550 | -10.32 |
| 586 | -10.42 |
| 623 | -10.51 |
| 660 | -10.61 |
| 696 | -10.71 |

## S11 CP2K example input file

A typical CP2K input file is given below. All geometries and CP2K inputs available on request.

```
&GLOBAL
  PRINT_LEVEL  MEDIUM
  PROJECT_NAME EXAMPLE
  RUN_TYPE  CELL_OPT
  FLUSH_SHOULD_FLUSH  T
 &END GLOBAL
 &MOTION
  &GEO_OPT
    TYPE  MINIMIZATION
    OPTIMIZER  LBFGS
    MAX_ITER  3000
    MAX_DR     2.9999999999999997E-04
    MAX_FORCE     4.5000000000000003E-05
    RMS_DR     1.4999999999999999E-04
    RMS_FORCE     3.0000000000000001E-05
    &CG
      MAX_STEEP_STEPS  0
      &LINE_SEARCH
        TYPE  2PNT
      &END LINE_SEARCH
    &END CG
  &END GEO_OPT
  &CELL_OPT
    OPTIMIZER  CG
    MAX_ITER  1000
    MAX_DR     3.0000000000000001E-03
    MAX_FORCE     4.4999999999999999E-04
    RMS_DR     1.5000000000000000E-03
    RMS_FORCE     2.9999999999999997E-04
    STEP_START_VAL  0
    TYPE  DIRECT_CELL_OPT
    EXTERNAL_PRESSURE     1.0000000000000000E+00
    KEEP_ANGLES  T
    PRESSURE_TOLERANCE     1.0000000000000000E+01
    &CG
      MAX_STEEP_STEPS  0
      RESTART_LIMIT     9.4999999999999996E-01
      &LINE_SEARCH
        TYPE  2PNT
      &END LINE_SEARCH
    &END CG
    &PRINT
      &PROGRAM_RUN_INFO  MEDIUM
      &END PROGRAM_RUN_INFO
      &CELL  MEDIUM
      &END CELL
    &END PRINT
  &END CELL_OPT
 &END MOTION
 &FORCE_EVAL
  METHOD  QS
  STRESS_TENSOR  ANALYTICAL
  &DFT
   BASIS_SET_FILE_NAME ./GTH_BASIS_SETS
   POTENTIAL_FILE_NAME ./POTENTIAL
    CHARGE  0
    &SCF
      MAX_SCF  20
      EPS_SCF     9.9999999999999995E-08
      SCF_GUESS  ATOMIC
      &OT  T
        MINIMIZER  DIIS
        PRECONDITIONER  FULL_ALL
        ENERGY_GAP     1.0000000000000000E-03
      &END OT
      &OUTER_SCF  T
        EPS_SCF     9.9999999999999995E-08
      &END OUTER_SCF
    &END SCF
    &QS
      EPS_DEFAULT     9.9999999999999998E-13
    &END QS
    &MGRID
      CUTOFF     6.5000000000000000E+02
    &END MGRID
    &XC
```

```
        DENSITY_CUTOFF     1.0000000000000000E-10
        GRADIENT_CUTOFF     1.0000000000000000E-10
        TAU_CUTOFF     1.0000000000000000E-10
        &XC_FUNCTIONAL  NO_SHORTCUT
          &PBE  T
            PARAMETRIZATION  ORIG
          &END PBE
        &END XC_FUNCTIONAL
      &END XC
      &POISSON
        POISSON_SOLVER  PERIODIC
        PERIODIC  XYZ
      &END POISSON
    &END DFT
    &SUBSYS
      &CELL
### ABC according to zeolite framework structure ###
        &CELL_REF
### ABCref 5% > ABC ###
        &END CELL_REF
      &END CELL
      &COORD
### zeolite crystal structure ###
      &END COORD
      &KIND Si
        BASIS_SET TZV2P-GTH
        POTENTIAL GTH-PBE-q4
      &END KIND
      &KIND O
        BASIS_SET TZV2P-GTH
        POTENTIAL GTH-PBE-q6
      &END KIND
      &KIND Al
        BASIS_SET TZV2P-GTH
        POTENTIAL GTH-PBE-q3
      &END KIND
      &KIND H
        BASIS_SET TZV2P-GTH
        POTENTIAL GTH-PBE-q1
      &END KIND
      &KIND Na
        BASIS_SET TZV2P-GTH
        POTENTIAL GTH-PBE-q9
      &END KIND
      &TOPOLOGY
        NUMBER_OF_ATOMS  ### according to zeolite framework structure
        MULTIPLE_UNIT_CELL  1 1 1
      &END TOPOLOGY
    &END SUBSYS
  &END FORCE_EVAL
```

**S12 Coarse-grain sifting for potential low-energy structures using GULP**

Investigating high-silica SSZ-13, 2 Al / unit cell, involved substituting a single uninodal Si T-site with Al and then sequentially substituting the remaining Si T-sites with a second Al, as described in the main body of the text. However, for lower silica frameworks, tending to a Si/Al of unity becomes increasingly more complex with each additional Al introduced to the framework. Introducing a third Al into the H-SSZ-13 framework results in 487, 344 possible combinations of 3 Al/ unit cell across the 36 T-sites, each with an associated proton at a neighbouring oxygen site. Manually constructing each of these models is rather time consuming, and fully optimising each model at the DFT level of theory is very compute intensive.

We attempted to reduce computational expense by utilising force field methods to determine potential low-energy structures, before fully optimising the most stable structures quantum mechanically using CP2K. Our method involved substituting the smaller 12 T-site, rhombohedral CHA unit-cell with a single aluminium, screening with GULP[10,11], and subsequently substituting the 50 lowest energy structures with a second aluminium. This methodology was repeated and continued until six of the 12 T-sites were occupied with Al (Si/Al = 1). We used both the Catlow potential and related modifications[12] – widely used in zeolite science, and the Clay Force Field (clayFF)[13], with a range of framework constraints, including constant pressure and constant volume calculations. However, disappointingly and despite extenisve re-fitting, we found only weak correlation (a coefficient of ~0.1) between the force field models and the DFT (PBE) data previously obtained for the high silica SSZ-13 case. Critically, the global minima differed and energy differences between configurations were in large absolute and relative error with respect to the DFT data. Despite extensive attempted optimisations of both forcefields, we were unable to obtain sufficient correlation between the force field and DFT data to allow us to use the forcefields as a pre-filter before DFT. In the light of this, we adopted the purely QM based method of stepwise substitution in the global minima structure described in the main body of the text.

**S13 References**

1. J. P. Perdew, K. Burke, M. Ernzerhof, *Phys. Rev. Lett.* **78**, 1396–1396 (1997).
2. Y. Zhang, W. Yang, *Phys. Rev. Lett.* **80**, 890–890 (1998).
3. A. D. Becke, *Phys. Rev. A*. **38**, 3098–3100 (1988).
4. C. Lee, W. Yang, R. G. Parr, *Phys. Rev. B*. **37**, 785–789 (1988).
5. S. J. Clark *et al.*, *Zeitschrift fuer Krist.* **220**, 567–570 (2005).
6. M. Guidon, J. Hutter, J. VandeVondele, *J. Chem. Theory Comput.* **6**, 2348–2364 (2010).
7. C. Adamo, V. Barone, *J. Chem. Phys.* **110**, 6158 (1999).
8. C. Baerlocher, L. . McCusker, D. H. Olson, *Atlas of Zeolite Framework Types* (sixth edit., 2007).
9. J. Klinowski, *Prog. Nucl. Magn. Reson. Spectrosc.* **16**, 237–309 (1984).
10. J. D. Gale, *J. Chem. Soc. Faraday Trans.* **93**, 629–637 (1997).
11. J. D. Gale, A. L. Rohl, *Mol. Simul.* **29**, 291–341 (2003).
12. M. J. Sanders, M. Leslie, C. R. A. Catlow, *J. Chem. Soc. Chem. Commun.* **19**, 1271 (1984).
13. R. T. Cygan, J.-J. Liang, A. G. Kalinichev, *J. Phys. Chem. B*. **108**, 1255–1266 (2004).